\documentclass[aps,prb,groupedaddress,twocolumn,showpacs,amsmath,amssymb]{revtex4-1}

\usepackage{stmaryrd}
\usepackage{times}
\usepackage{graphicx}% Include figure files
\usepackage{dcolumn}% Align table columns on decimal point
\usepackage{bm}% bold math
\usepackage{subeqnarray}
\usepackage{cases}
\usepackage{upgreek}
\begin{document}

%\preprint{}

\title{Electrical control over perpendicular magnetization switching driven by spin-orbit torques}


\author{X. Zhang}
\affiliation{Institute of Physics, Chinese Academy of Sciences, Beijing National Laboratory for Condense Matter Physics, Beijing, 100190, China}

\author{C. H. Wan}
\email[Email: ]{wancaihua@iphy.ac.cn}
\affiliation{Institute of Physics, Chinese Academy of Sciences, Beijing National Laboratory for Condense Matter Physics, Beijing, 100190, China}

\author{Z. H. Yuan}
\affiliation{Institute of Physics, Chinese Academy of Sciences, Beijing National Laboratory for Condense Matter Physics, Beijing, 100190, China}

\author{Q. T. Zhang}
\affiliation{Institute of Physics, Chinese Academy of Sciences, Beijing National Laboratory for Condense Matter Physics, Beijing, 100190, China}

\author{H. Wu}
\affiliation{Institute of Physics, Chinese Academy of Sciences, Beijing National Laboratory for Condense Matter Physics, Beijing, 100190, China}

\author{L. Huang}
\affiliation{Institute of Physics, Chinese Academy of Sciences, Beijing National Laboratory for Condense Matter Physics, Beijing, 100190, China}

\author{W. J. Kong}
\affiliation{Institute of Physics, Chinese Academy of Sciences, Beijing National Laboratory for Condense Matter Physics, Beijing, 100190, China}

\author{C. Fang}
\affiliation{Institute of Physics, Chinese Academy of Sciences, Beijing National Laboratory for Condense Matter Physics, Beijing, 100190, China}

\author{U. Khan}
\affiliation{Institute of Physics, Chinese Academy of Sciences, Beijing National Laboratory for Condense Matter Physics, Beijing, 100190, China}

\author{X. F. Han}
\email[Email: ]{xfhan@iphy.ac.cn}
\affiliation{Institute of Physics, Chinese Academy of Sciences, Beijing National Laboratory for Condense Matter Physics, Beijing, 100190, China}
%\thanks{}


\date{\today}

\begin{abstract}
Flexible control of magnetization switching by electrical manners is crucial for applications of spin-orbitronics. Besides of a switching current that is parallel to an applied field, a bias current that is normal to the switching current is introduced to tune the magnitude of effective damping-like and field-like torques and further to electrically control magnetization switching. Symmetrical and asymmetrical control over the critical switching current by the bias current with opposite polarities is both realized in Pt/Co/MgO and $\alpha$-Ta/CoFeB/MgO systems, respectively. This research not only identifies the influences of field-like and damping-like torques on switching process but also demonstrates an electrical method to control it.
\end{abstract}


\maketitle

\section{INTRODUCTION}
Spin-orbitronics\cite{Kuschel2014,Manchon2014}, aiming at current or voltage control of magnetization (\textbf{M}) via spin-orbit coupling (SOC) effect, has gradually manifested itself charming prospect in nonvolatile magnetic storage and programmable spin-logic applications. Spin Hall effect (SHE) in heavy metals\cite{Yakonov1971,Hirsch1999,Zhang2000,Sinova2015} or topologic insulators\cite{Fan2014} and Rashba effect\cite{Rashba1984,MihaiMiron2010} at the heavy metal/ferromagnetic metal interfaces are two broadly utilized effects to realize spin-orbitronics due to their large SOC strength. With the aid of magnetic field, SHE induced magnetization switching has already been realized in many systems comprising a magnetic layer (Co, CoFeB, NiFe) sandwiched by an oxide layer(AlO$_{\mbox{\footnotesize \emph{x}}}$, MgO) and a heavy metal layer (Pt, $\beta$-Ta, W) with not only in-plane anisotropy\cite{Fan2013,Liu2012} but perpendicular anisotropy\cite{Miron2011,LiuPRL2012,Pai2012,Qiu2015}. Recently, field-free magnetization switching via current has been also achieved in a wedged Ta/CoFeB/TaO$_{\mbox{\footnotesize \emph{x}}}$\cite{Yu2014} or antiferromagnetic/ferromagnetic coupled perpendicular systems\cite{Fukami2016,vandenBrink2016,Lau2015}.

In those perpendicular systems, current can generate via SHE effect a damping-like torque which balances effective torques from perpendicular anisotropy and in-plane bias field and consequently switches magnetization as it becomes large enough. In these previous researches, mainly spin Hall torque (along \emph{x} axis) induced by one current (namely, switching current \emph{I} along \emph{y} axis) applied along the direction of an applied or effective magnetic field is taken into account while the influence of field-like torque (along \emph{y} axis) on magnetic reversal process is rarely experimentally testified. Definition of coordinates is shown in Fig. \ref{fig1}(a).

Here, we introduced another current (namely, bias current $I_{\mbox{\scriptsize B}}$ along \emph{x} axis) to electrically control magnetization switching process (Fig. \ref{fig1}(a)). The damping-like and field-like torques of the bias current have the same symmetry with the field-like and damping-like torque of the switching current, respectively. Therefore, as shown below, the influences of both field-like torque and damping-like torque of the switching current on magnetization switching process become visible with tuning the magnitude of the bias current. Furthermore, the main features of aforementioned results can be well reproduced by a macrospin model which provides further understanding. This work can not only help to distill the influences of different kinds of torques on the switching process but also demonstrate a practical manner of controlling SHE-driven magnetization switching process by electrically tuning the magnitude of effective damping-like and field-like torques.

\section{EXPERIMENTAIL METHOD}
SiO$_{2}$//Ta(5)/Co$_{20}$Fe$_{60}$B$_{20}$(1)/MgO(2)/Pt(3) and SiO$_{2}$//Pt(5)/Co(0.8)/MgO(2)/Pt(3) (thickness in nanometer) stacks were provided by Singulus GmbH. They were magnetron-sputtered at room temperature. They have intrinsically in-plane anisotropy. After annealing at 400$^{\circ}\mathrm{C}$ and 10$^{-3}$ Pa for 1 h in a perpendicular field of 0.7 T could the stacks exhibit strong PMA. Raw films were then patterned by ultraviolet lithography and the following two-step argon ion etching into Hall bars with the size of the center squares being 20 $\mu$m (Fig. \ref{fig1}(a)). Cu(10 nm)/Au(30 nm) electrodes were finally deposited to make contacts with four legs of Hall bars. After device microfabrication, the Hall bars were measured with two Keithley 2400 sourcemeters and Keithley 2182 voltmeter sourcing devices and measuring Hall voltages, respectively. Meanwhile, PPMS-9T (Quantum Design) provided magnetic fields with proper directions. The two Keithley 2400 sourcemeters first provided the current pulses to the Hall bar. One applied switching current along the \emph{y} axis and the other applied bias current along the \emph{x} axis to the sample with a duration time of 50 ms. Then the two Keithley 2400 stopped sourcing after the duration time. After waiting for 100 ms, one Keithley 2400 applied another current pulse of 1 mA along the \emph{y} axis to the sample for 100 ms. At the end of this pulse, Keithley 2182 picked up the Hall voltage along the \emph{x} axis. Then the Keithley 2400 was switched off. After 100 ms, the next round of destabilizing-measuring process was performed.
\section{RESSULTS AND DISCUSSION}
\subsection{Experiment}
\begin{figure}
  \centering
  \includegraphics[width=0.45\textwidth]{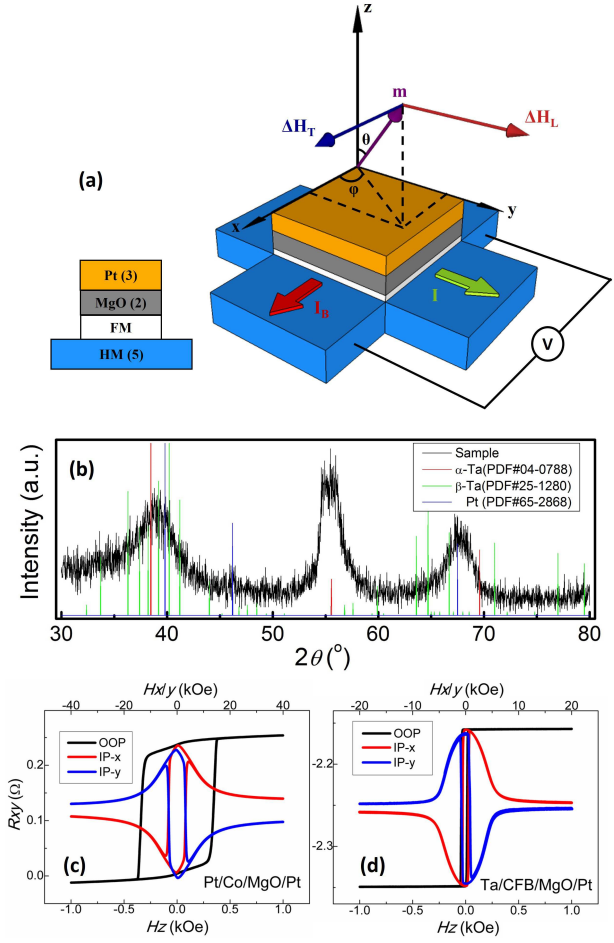}\\
  \caption{(color online). (a) Sample structure of a Hall bar. (b) Glancing XRD pattern of Ta/Co$_{20}$Fe$_{60}$B$_{20}$/MgO/Pt stacks. (c) and (d) show \emph{H}-dependence of Hall resistance of the PCM and TFM, respectively, as field is along \emph{x}/\emph{y}/\emph{z} axis. The Hall resistance $R_{\mbox{\footnotesize \emph{xy}}}\equiv V_{\mbox{\footnotesize \emph{x}}}/I_{\mbox{\footnotesize \emph{y}}}$ in (c) and (d) is obtained as $I_{\mbox{\footnotesize \emph{y}}}$=1 mA and $I_{\mbox{\scriptsize B\emph{x}}}$=0.}\label{fig1}
\end{figure}

Two typical perpendicular systems Sub//Pt(5)/Co(0.8)/MgO(2)/Pt(3) (PCM for short) and Sub//Ta(5)/Co$_{20}$Fe$_{60}$B$_{20}$(1.0)/MgO(2)/Pt(3) (TFM) are used for comparison. Thickness is in nanometers. Here Ta is in $\alpha$-phase instead of $\beta$-phase (Fig. \ref{fig1}(b)). The strong peak at 2$\theta$=55.6$^{\circ}$ can be only ascribed to (200) plane of $\alpha$-Ta. Absence of the two main peaks at 2$\theta$=63.6$^{\circ}$ and 64.7$^{\circ}$ corresponding to (631) and (413) planes of $\beta$-phase, respectively, indicates nonexistence of $\beta$-phase. The wide peak at 2$\theta$=39$^{\circ}$ can be attributed to the merge of (110) plane of $\alpha$-Ta and (111) plane of Pt. The other wide peak at 2$\theta$=68$^{\circ}$ can be due to the merge of (211) plane of $\alpha$-Ta and (220) plane of Pt.

$M_{0}t$ of PCM and TFM measured by Vibration Sample Magnetometry is 125 $\mu$emu/cm$^{2}$ and 145 $\mu$emu/cm$^{2}$, respectively. $M_{0}$ and $t$ is saturated magnetization and thickness of magnetic layer, respectively. Hall measurement demonstrates perpendicular magnetic anisotropy (PMA) of both systems. PCM shows higher PMA energy than TFM. Anisotropy field ($H_{\mbox{\footnotesize an}}$) of PCM and TFM is about 13.6 kOe and 5.8 kOe, respectively (Figs. \ref{fig1}(c) and \ref{fig1}(d)). Sophisticated harmonic lock-in technique\cite{Pi2010,Kim2012} is applied here to characterize spin-orbit torques of the above systems induced by applied current. The effective longitudinal field $\Delta$$H_{\mbox{\scriptsize L}}$ and effective transverse field $\Delta$$H_{\mbox{\scriptsize T}}$ corresponding to damping-like torque and field-like torque, respectively, are shown in Fig. \ref{fig1}. Sample structure is also shown (Fig. \ref{fig1}(a)).

\begin{figure}
  \centering
  \includegraphics[width=0.45\textwidth]{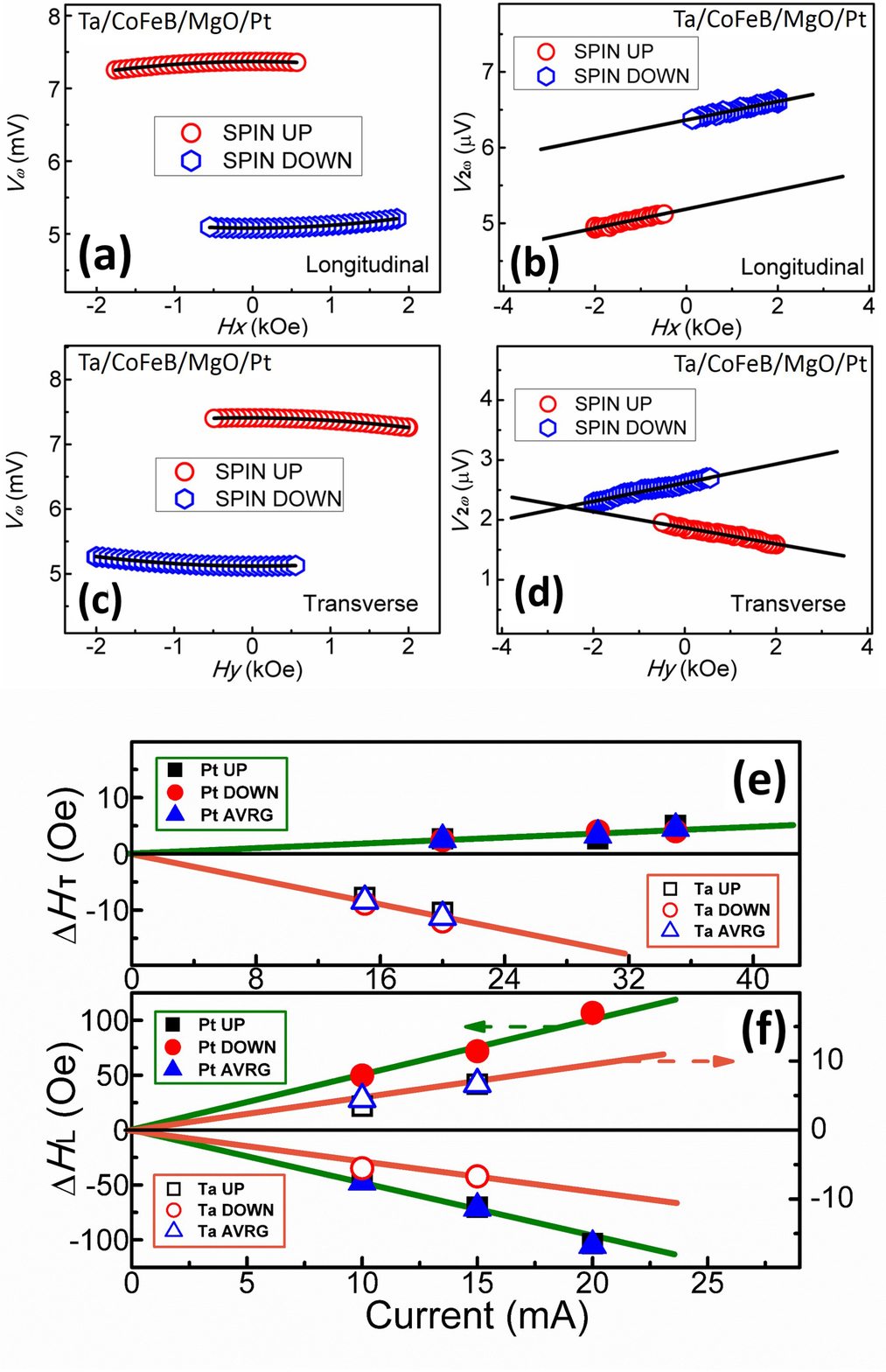}\\
  \caption{(color online). The $H_{\mbox{\footnotesize \emph{y}}}$ dependence of (a) $V^\omega$ and (b) $V^{2\omega}$ and the $H_{\mbox{\footnotesize \emph{x}}}$ dependence of (c) $V^\omega$ and (d) $V^{2\omega}$ in TFM. (e) and (f) shows, respectively, the current dependence of $\Delta H_{\mbox{\scriptsize T}}$ and $\Delta H_{\mbox{\scriptsize L}}$ in both TFM and PCM films. Their linear fittings with zero intercept are also shown. As measuring the effective fields induced by the switching current, we applied no bias current. FM and HM denote ferromagnetic and heavy metal, respectively.}\label{fig2}
\end{figure}

During measurement, current density ($j_{\mbox{\footnotesize \emph{y}}}$=$j_{\mbox{\footnotesize \emph{y}0}}\sin{\omega t}$) is applied along +\emph{y} axis. Magnetic field (\emph{H}) is applied along \emph{x} or \emph{y} axis. Direction of \emph{H} determines which torque can be detected. $H_{\mbox{\footnotesize \emph{x}}}$ and $H_{\mbox{\footnotesize \emph{y}}}$ are respectively used to measure current-induced field-like torque (or effective transverse field $\Delta H_{\mbox{\scriptsize T}}$ corresponding to the field-like torque) and damping-like torque (or effective longitudinal field $\Delta H_{\mbox{\scriptsize L}}$ corresponding to the damping-like torque). First and second harmonic Hall voltages along \emph{x} axis
($V^{\omega}_{\mbox{\footnotesize \emph{x}}}$=$V^{\omega}_{\mbox{\footnotesize \emph{x}0}}\sin{\omega t}$ and $V^{2\omega}_{\mbox{\footnotesize \emph{x}}}$=$V^{2\omega}_{\mbox{\footnotesize \emph{x}0}}\cos{2\omega t}$) are picked up to indirectly show direction of magnetization (\textbf{M}) respective to the +\emph{z} axis and $j_{\mbox{\footnotesize \emph{y}}}$-tuned \textbf{M} change, accordingly. $V^{\omega}_{\mbox{\footnotesize \emph{x}0}}$ and $V^{2\omega}_{\mbox{\footnotesize \emph{x}0}}$ exhibit parabolic and linear field dependence as \textbf{M} around $\pm$\emph{z}, respectively. Especially, the $V^{2\omega}_{\mbox{\footnotesize 0}}$ vs. \emph{H} curves (Fig. \ref{fig2}(b)) exhibit the same slopes at $\pm m_{\mbox{\footnotesize \emph{z}}}$ as \emph{H} is along \emph{y} while they exhibit opposite slopes as \emph{H} is along \emph{x} (Fig. \ref{fig2}(d)). From the slopes as well as ${\partial^2 V^{\omega}}/{\partial H^2}$ (Figs. \ref{fig2}(a) and \ref{fig2}(c)) can we obtain $\Delta$$H_{\mbox{\scriptsize L}}$ along \emph{y} axis and $\Delta$$H_{\mbox{\scriptsize T}}$ along \emph{x} axis via $\Delta$$H_{\mbox{\scriptsize L/T}}$=-2($\partial V^{2\omega}/{\partial H_{\mbox{\footnotesize \emph{y}/\emph{x}}}}$)$/$(${\partial^2 V^{\omega}}/{\partial H^2_{\mbox{\footnotesize \emph{y}/\emph{x}}}}$). Here $\Delta$\textbf{H}$_{\mbox{\scriptsize L}}$ parallel to \bm{$\sigma$}$\times$\textbf{M} originates from spin Hall effect. $\Delta$\textbf{H}$_{\mbox{\scriptsize T}}$ parallel to \bm{$\sigma$} originates from Rashba field as well as Ostered field. The \bm{$\sigma$} is the spin current density induced by the \textbf{j}$_{\mbox{\footnotesize \emph{y}}}$ via \bm{$\sigma$}$\propto$ \textbf{j}$_{\mbox{\footnotesize \emph{y}}}$$\times$ \textbf{z}.

The $\Delta$$H_{\mbox{\scriptsize L/T}}$ shows linear dependence on applied current density $j_{\mbox{\footnotesize \emph{y}}}$ with zero intercepts as expected. Parameter $\beta_{\mbox{\scriptsize L/T}}$ defined as $d\Delta H_{\mbox{\scriptsize L/T}}/dj_{\mbox{\footnotesize \emph{y}}}$ characterizes conversion efficiency from charge current to effective field. Here, $j_{\mbox{\footnotesize \emph{y}}}$=$I/(wh_{\mbox{\tiny HM}})$. \emph{I} is the switching current, \emph{w} is the width of Hall bar (20 $\mu$m) and $h_{\mbox{\tiny HM}}$ is the thickness of the heavy metal (5 nm). 1 mA of \emph{I} thus corresponds to 1 MA/cm$^{2}$ of $j_{\mbox{\footnotesize \emph{y}}}$. The shielding effect of the ferromagnetic layer and anti-oxidation layer is ignored. Thus, $j_{\mbox{\footnotesize \emph{y}}}$ and $\beta_{\mbox{\scriptsize L/T}}$ should be deemed as an upper and lower bound, respectively. The $\beta_{\mbox{\scriptsize L}}$ is about -40 nm and +4 nm for PCM and TFM, respectively (Fig. \ref{fig2}(e)). Meanwhile, the $\beta_{\mbox{\scriptsize T}}$ is about +1.2 nm and -4 nm for PCM and TFM, respectively (Fig. \ref{fig2}(f)). Especially, $\beta_{\mbox{\scriptsize L/T}}$ of $\alpha$-Ta and Pt has opposite signs. The $\beta_{\mbox{\scriptsize L/T}}$ of Pt is reported in the order of 1 $\mu$m-1 nm\cite{MihaiMiron2010,Fan2013,LiuPRL2012,Pi2010,XinFan2014} in different systems. Our value is closer to that of Liu\cite{LiuPRL2012} and Fan\cite{Fan2013}. Besides, $|\beta_{\mbox{\scriptsize L, Pt}}|\gg|\beta_{\mbox{\scriptsize T, Pt}}|$, consistent with the results of Liu\cite{LiuPRL2012}. The $\beta_{\mbox{\scriptsize L/T}}$ of Ta in Ta/CoFeB/MgO system is thoroughly researched by Kim\cite{Kim2012}. It is in the order of 2-20 nm, depending on thickness of Ta and CoFeB. Besides, their results show $\beta_{\mbox{\scriptsize T, Ta}}$ can be comparable and even larger than $\beta_{\mbox{\scriptsize L, Ta}}$. Our measured values are within their range and $|\beta_{\mbox{\scriptsize L, Ta}}|$ is equal to $|\beta_{\mbox{\scriptsize T, Ta}}|$. However, the $\beta_{\mbox{\scriptsize L}}$ of $\alpha$-Ta is smaller than that of $\beta$-Ta\cite{Kim2012}. Ratio of $\beta_{\mbox{\scriptsize T}}$/$\beta_{\mbox{\scriptsize L}}$ for Pt and $\alpha$-Ta is -0.03 and -1, respectively. Field-like torque can be nearly neglected in the PCM while it cannot be ignored in the TFM, which provides us a couple of ideal systems to research the influence of field-like torque and damping-like torque on switching behavior of perpendicular films. The reason why field-like torque is insignificant and significant in PCM and TFM system respectively, we think, is that the two systems may have different interfacial potentials due to different work functions of Pt (5.3 eV), Co (4.4 eV), Fe (4.3 eV), and Ta (4.1 eV)\cite{Michaelson1977,Skriver1992} as elaborated in Ref. [\onlinecite{Barnes2014}].
\begin{figure*}
  \centering
  \includegraphics[width=0.8\textwidth]{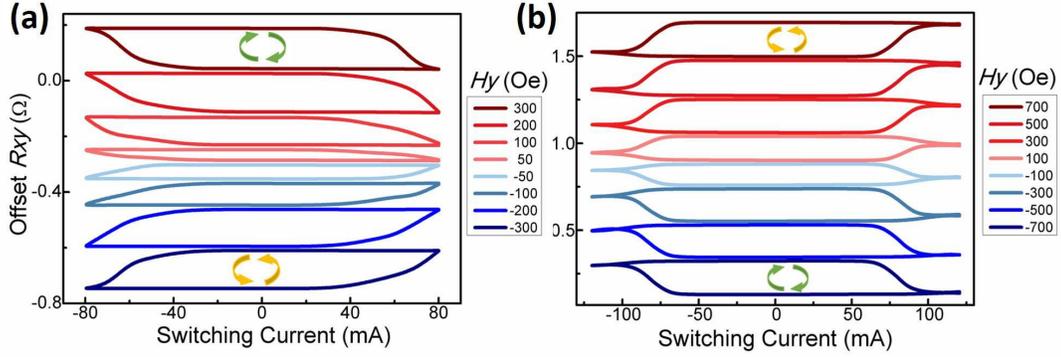}\\
  \caption{(color online). The dependence of $R_{\mbox{\footnotesize \emph{xy}}}$ on switching current (\emph{I}) in (a) TFM and (b) PCM systems under different $H_{\mbox{\footnotesize \emph{y}}}$.}\label{fig3}
\end{figure*}

In the following, we will use the PCM with $\beta_{\mbox{\scriptsize T}}$/$\beta_{\mbox{\scriptsize L}}$=-0.03 and the TFM with $\beta_{\mbox{\scriptsize T}}$/$\beta_{\mbox{\scriptsize L}}$=-1 to study the influence of $I_{\mbox{\scriptsize B}}$ on switching behaviors and introduce the underneath mechanism based on a macrospin model. \emph{I} and $H_{\mbox{\footnotesize \emph{y}}}$ are applied along \emph{y}. $I_{\mbox{\scriptsize B}}$ is applied along \emph{x}. As $I_{\mbox{\scriptsize B}}$=0, \textbf{M} can be switched back and forth between spin-up state and spin-down state (Fig. \ref{fig3}) by scanning \emph{I} under nonzero $H_{\mbox{\footnotesize \emph{y}}}$. Due to opposite spin Hall angle, switching direction is opposite for PCM and TFM with the same measurement setup. For example, switching direction for TFM and PCM is clockwise and anticlockwise, respectively, at positive $H_{\mbox{\footnotesize \emph{y}}}$. Sign reversal of $H_{\mbox{\footnotesize \emph{y}}}$ leads to reversal of the switching direction. Fig. \ref{fig3} also shows nearly a full magnetization switching can be realized as $H_{\mbox{\footnotesize \emph{y}}}$=0.3 kOe for TFM. In this condition, critical switching current ($I_{\mbox{\scriptsize C}}$) is 63.5 mA. Meanwhile, the $I_{\mbox{\scriptsize C}}$ for PCM is about 80 mA as $H_{\mbox{\footnotesize \emph{y}}}$=0.7 kOe. These results manifest $\alpha$-Ta can also function as a high efficient converter from charge current to spin current besides of Pt and $\beta$-Ta.
\begin{figure*}
  \centering
  \includegraphics[width=0.8\textwidth]{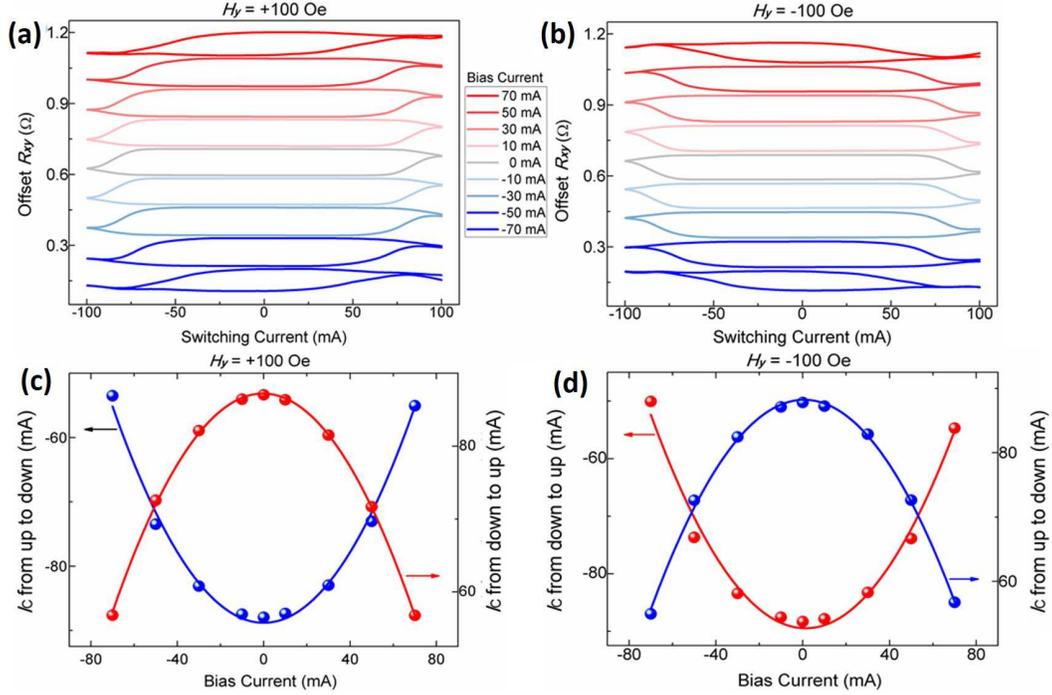}\\
  \caption{(color online). The switching current dependence of $R_{\mbox{\footnotesize \emph{xy}}}$ of the PCM under different bias current as (a) $H_{\mbox{\footnotesize \emph{y}}}$=100 Oe and (b) -100 Oe and the dependence of $I_{\mbox{\scriptsize C}}$ on $I_{\mbox{\scriptsize B}}$ as (c) $H_{\mbox{\footnotesize \emph{y}}}$=100 Oe and (d) -100 Oe. Red and blue dots in (c) and (d) show, respectively, the $I_{\mbox{\scriptsize C}}$ of transitions from down-state to up-state and from up-state to down-state. The dependence of $I_{\mbox{\scriptsize C}}$ on $I_{\mbox{\scriptsize B}}$ in (c) and (d) could be well reproduced by parabolic fittings.}\label{fig4}
\end{figure*}

As shown in Figs. \ref{fig4}(a) and \ref{fig4}(b) elevated $I_{\mbox{\scriptsize B}}$ can significantly reduce the $I_{\mbox{\scriptsize C}}$ in the PCM system. For example, $I_{\mbox{\scriptsize C}}$=88 mA as $I_{\mbox{\scriptsize B}}$=0 mA while $I_{\mbox{\scriptsize C}}$=73 mA as $I_{\mbox{\scriptsize B}}$=50 mA. $I_{\mbox{\scriptsize C}}$ decreases by 17\%. Meanwhile, positive and negative $I_{\mbox{\scriptsize B}}$ leads to nearly the same amount of reduction, no matter the sign of $H_{\mbox{\footnotesize \emph{y}}}$ as shown by the parabolic fitting lines in Figs. \ref{fig4}(c) and \ref{fig4}(d). This $I_{\mbox{\scriptsize B}}$-induced decrease in $I_{\mbox{\scriptsize C}}$ can be ascribed to the damping-like torque from $I_{\mbox{\scriptsize B}}$ as shown in the theoretical part below. It is worthy of accentuating that the damping-like torque of $I_{\mbox{\scriptsize B}}$ shares the similar symmetry with the field-like torque of \emph{I} and thus a large field-like torque of \emph{I} could also in principle reduce the $I_{\mbox{\scriptsize C}}$.

Certainly, $I_{\mbox{\scriptsize B}}$ will heat magnetic films as well and in principle reduce the effective $H_{\mbox{\footnotesize an}}$, which could also reduce $I_{\mbox{\scriptsize C}}$. However, our experiment shows that $I_{\mbox{\scriptsize C}}$ varies little as changing duration time of $I_{\mbox{\scriptsize B}}$ from 50 ms to 1 s, which indicates that thermal effect is at least not dominating factor in determining $I_{\mbox{\scriptsize C}}$ here.

\begin{figure*}
  \centering
  \includegraphics[width=0.8\textwidth]{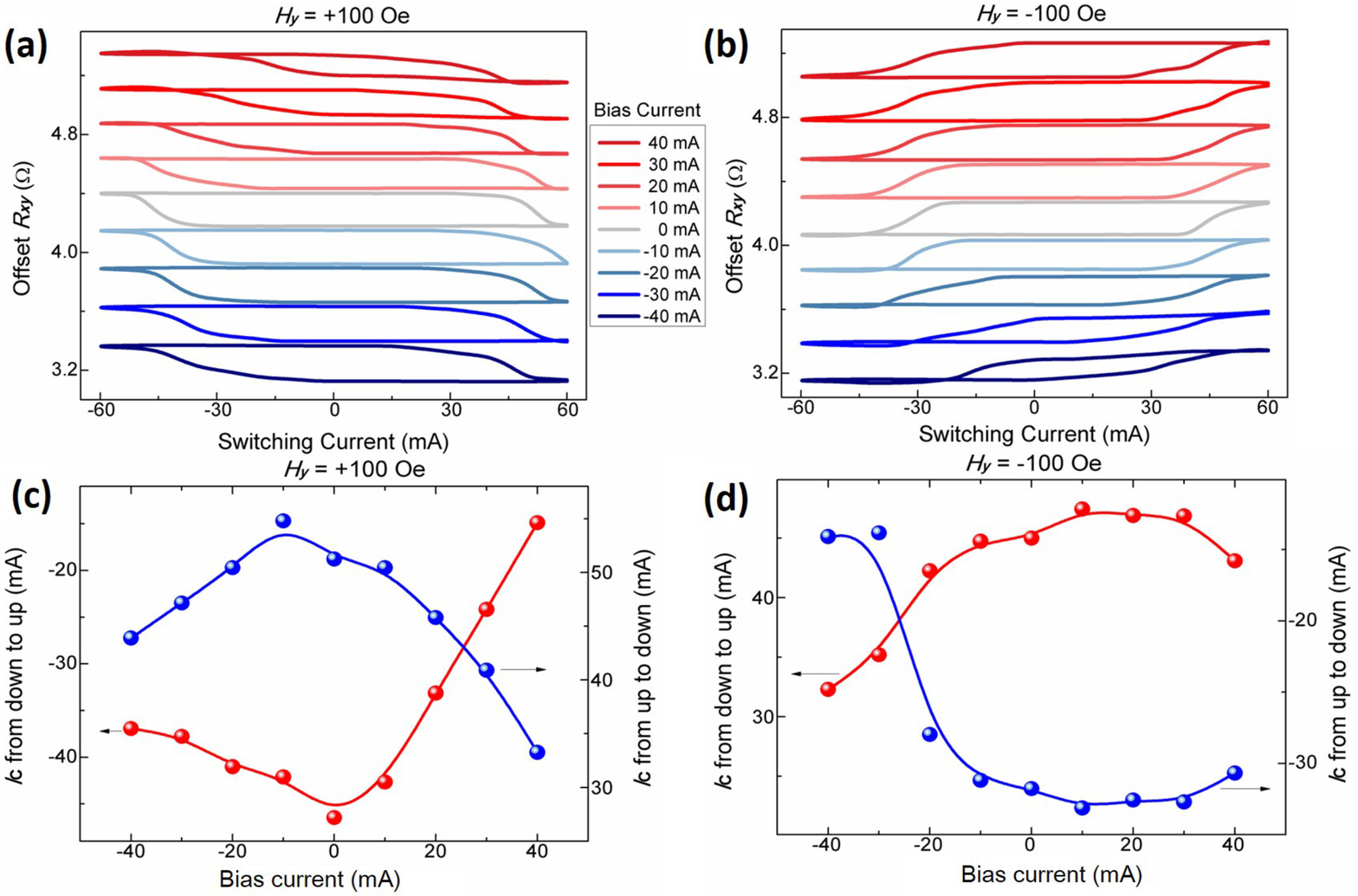}\\
  \caption{(color online). The switching current dependence of $R_{\mbox{\footnotesize \emph{xy}}}$ of the TFM under different bias current as (a) $H_{\mbox{\footnotesize \emph{y}}}$=+100 Oe and (b) -100 Oe and the dependence of $I_{\mbox{\scriptsize C}}$ on $I_{\mbox{\scriptsize B}}$ as (c) $H_{\mbox{\footnotesize \emph{y}}}$=+100 Oe and (d) -100 Oe.}\label{fig5}
\end{figure*}

On the other hand, TFM system manifests a different response to $I_{\mbox{\scriptsize B}}$ with different symmetry in comparison with the PCM counterpart. As shown in Figs. \ref{fig5}(a) and \ref{fig5}(c) for $H_{\mbox{\footnotesize \emph{y}}}$=+100 Oe and the transition from down-state to up-state, $I_{\mbox{\scriptsize C}}$ is reduced by about 67\% under $I_{\mbox{\scriptsize B}}$=40 mA while it is only reduced by 20\% under $I_{\mbox{\scriptsize B}}$=-40 mA. In contrast, for $H_{\mbox{\footnotesize \emph{y}}}$=-100 Oe and the transition from up-state to down-state (Figs. \ref{fig5}(b) and \ref{fig5}(d)), besides of the opposite switching direction, the effect of $I_{\mbox{\scriptsize B}}$ on $I_{\mbox{\scriptsize C}}$ is also reversed, i.e. $I_{\mbox{\scriptsize C}}$ decreased only by about 5\% under $I_{\mbox{\scriptsize B}}$=40 mA while it decreased remarkably by 53\% under $I_{\mbox{\scriptsize B}}$=-40 mA. Here, the asymmetric response of $I_{\mbox{\scriptsize C}}$ to positive and negative $I_{\mbox{\scriptsize B}}$ cannot be interpreted by damping-like torque induced by $I_{\mbox{\scriptsize B}}$ or heating effect as shown in the case of PCM. Instead, field-like torque of $I_{\mbox{\scriptsize B}}$ is a key contributor to the asymmetry as shown below.

\subsection{Macrospin model}
In order to interpret the different response of PCM and TFM to $I_{\mbox{\scriptsize B}}$, we have turned to a macrospin model (more details in Appendix). The magnetic energy includes uniaxial anisotropy energy $K\sin{2\theta}$ and Zeeman energy $-H_{\mbox{\footnotesize \emph{y}}}M_0\sin\theta\sin\varphi$ where $\theta$ and $\varphi$ is the polar angle between \textbf{M} and the +\emph{z} axis and the azimuthal angle between in-plane projection of \textbf{M} and the +\emph{x} axis, respectively (Fig. \ref{fig1}(a)). \emph{I} and $I_{\mbox{\scriptsize B}}$ provide both a damping-like torque and a field-like torque on \textbf{M} with efficiency characterized by $\beta_{\mbox{\scriptsize T}}$/$\beta_{\mbox{\scriptsize L}}$. We use parameter \emph{a} in unit of $H_{\mbox{\footnotesize an}}\equiv2K/\mu_0M_0$ to denote the damping-like torque provided by \emph{I}, parameter \emph{c} to denote the ratio of $I_{\mbox{\scriptsize B}}/I$ and parameter \emph{b} to denote the ratio of $\beta_{\mbox{\scriptsize T}}$/$\beta_{\mbox{\scriptsize L}}$. Actually, \emph{c} reflects the angle of total current density with respect to the direction of magnetic field. As \emph{I} and $I_{\mbox{\scriptsize B}}$ are both applied, torque equilibrium condition requires satisfaction of Eq. (\ref{torque}).
%\begin{widetext}
\begin{equation}\label{torque}
\begin{aligned}
  0=&\vec{m}\times \vec{H}_{eff}\\
  &+a\vec{m}\times(-\hat{e}_x)\times \vec{m}+ab\vec{m}\times(-\hat{e}_x)\\
  &+ac\vec{m}\times \hat{e}_y \times \vec{m}+abc\vec{m}\times \hat{e}_y
\end{aligned}
\end{equation}
%\end{widetext}
Here \textbf{H}$_{\mbox{\footnotesize eff}}$=$-\nabla_{\mbox{\scriptsize \emph{M}}}$\emph{E}, \textbf{m}$\equiv$\textbf{M}/$M_0$, $E\equiv K\sin^2\theta-\mu_0M_0H_{\mbox{\footnotesize \emph{y}}}\sin\theta\cos\varphi$, \textbf{\emph{e}}$_{\mbox{\footnotesize \emph{x}}}$ and \textbf{\emph{e}}$_{\mbox{\footnotesize \emph{y}}}$ is a unit vector along the \emph{x} and \emph{y} axes, respectively. The 2$^{\mbox{\footnotesize nd}}$ and 3$^{\mbox{\footnotesize rd}}$ term in the RHS of Equation (\ref{torque}) is damping-like and field-like torque from \emph{I} while the 4$^{\mbox{\footnotesize th}}$ and 5$^{\mbox{\footnotesize th}}$ term is damping-like and field-like torque from $I_{\mbox{\scriptsize B}}$, respectively. Equation (\ref{torque}) can be further reduced as scalar equations. Equation (\ref{simplified}) is one of them.
\begin{widetext}
\begin{equation}\label{simplified}
%\begin{aligned}
  \sin\theta\cos\theta-\frac{[a^2(b^2+1)(c^2+1)+h^2_y-2abch_y]}{h_y-a\cos\theta-abc}\cos\theta\sin\varphi+\frac{ah_y(1+\cos^2\theta)}{h_y-a\cos\theta-abc}\sin\varphi=0
%\end{aligned}
\end{equation}
\end{widetext}

If $I_{\mbox{\scriptsize B}}$=0 and \emph{b}=0, $\sin\theta\cos\theta-h_{\mbox{\footnotesize \emph{y}}}\cos\theta\sin\varphi+a\sin\varphi=0$, which shares the similar form as derived by Liu\cite{LiuPRL2012} and Yan\cite{Yan2015}. Here $h_{\mbox{\footnotesize \emph{y}}}\equiv H_{\mbox{\footnotesize \emph{y}}}/H_{\mbox{\footnotesize an}}$. Comparing Eq. (\ref{simplified}) with the simplified one as $I_{\mbox{\scriptsize B}}$=0 and \emph{b}=0, we can see that introduction of $I_{\mbox{\scriptsize B}}$ leads to an effective $h_{\mbox{\footnotesize \emph{y}}}^{\mbox{\footnotesize eff}}$ and an effective damping-like torque $a^{\mbox{\footnotesize eff}}$ as expressed in Equation (\ref{coefficient}).
%\begin{widetext}
\begin{subequations}\label{coefficient}
\begin{equation}
  h_y^{eff}=\frac{[a^2(b^2+1)(c^2+1)+h^2_y-2abch_y]}{h_y-a\cos\theta-abc}
\end{equation}
\begin{equation}
  a^{eff}=\frac{ah_y(1+\cos^2\theta)}{h_y-a\cos\theta-abc}
\end{equation}
\end{subequations}
%\end{widetext}

\begin{figure}
  \centering
  \includegraphics[width=0.45\textwidth]{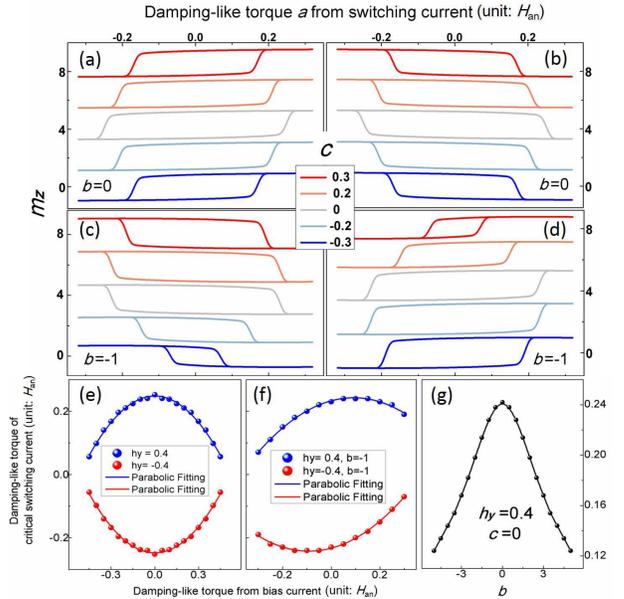}\\
  \caption{(color online). Dependence of $m_{\mbox{\footnotesize \emph{z}}}$ on damping-like torque of switching current \emph{a} (in unit of $H_{\mbox{\footnotesize an}}$) for different \emph{c}, as (a) $h_{\mbox{\footnotesize \emph{y}}}$=0.4, (b) $h_{\mbox{\footnotesize \emph{y}}}$=-0.4 with \emph{b}=0 and (c) $h_{\mbox{\footnotesize \emph{y}}}$=0.4, (d) $h_{\mbox{\footnotesize \emph{y}}}$=-0.4 with \emph{b}=-1. (e) and (f) $\tau_{\mbox{\footnotesize c}}$ as a function of damping-like torque of $I_{\mbox{\scriptsize B}}$ under $h_{\mbox{\footnotesize \emph{y}}}$=$\pm$0.4 as \emph{b}=0 and \emph{b}=-1, respectively. Here $\tau_{\mbox{\footnotesize c}}$ is obtained by the transition from spin-down to spin-up state. (g) $\tau_{\mbox{\footnotesize c}}$ as a function of \emph{b} as \emph{c}=0 and $h_{\mbox{\footnotesize \emph{y}}}$=0.4.}\label{fig6}
\end{figure}

Simulated results according to Eq. (\ref{torque}) are shown in Fig. \ref{fig6} where $\tau_{\mbox{\tiny C}}\propto$$I_{\mbox{\scriptsize C}}$ is the critical damping-like torque of \emph{I}. As \emph{c}=0, a nonzero \emph{b} can significantly reduce critical switching current ($I_{\mbox{\scriptsize C}}$), regardless of its sign (Fig. \ref{fig6}(g)). $I_{\mbox{\scriptsize C}}$ decreases by 5.8\% and 42\% as \emph{b}=$\pm$1 and \emph{b}=$\pm$3.6\cite{Kim2012}, respectively, compared with the $I_{\mbox{\scriptsize C}}$ as \emph{b}=0. This trend is consistent with the result in the PCM sample in which the damping-like torque of $I_{\mbox{\scriptsize B}}$ can mimic the influence of the field-like torque of \emph{I}. Though it cannot reverse \textbf{M} directly, large Rashba effect can still help to effectively reduce $I_{\mbox{\scriptsize C}}$.

As \emph{b}=0, bias current (\emph{c}$\neq$0) can notably decrease $I_{\mbox{\scriptsize C}}$ and the amount of the reduction in $I_{\mbox{\scriptsize C}}$ does not depend on polarity of \emph{c} (Figs. \ref{fig6}(a) and \ref{fig6}(b)), which manifests similar characteristics with the switching behaviors of the PCM sample. As \emph{b}=-1 and $h_{\mbox{\footnotesize \emph{y}}}$=0.4 (Fig. \ref{fig6}(c)), \emph{c}=0.3 and \emph{c}=-0.3 will result in asymmetric decrease in $I_{\mbox{\scriptsize C}}$. Here \emph{c}=-0.3 is more effective in reducing $I_{\mbox{\scriptsize C}}$. However, as $h_{\mbox{\footnotesize \emph{y}}}$=-0.4 (Fig. \ref{fig6}(d)), $I_{\mbox{\scriptsize C}}$ reduces more in the case of \emph{c}=+0.3. These characteristics (Figs. \ref{fig6}(c) and \ref{fig6}(d)) well reproduce the results of the TFM sample in Figs. \ref{fig5}(a) and \ref{fig5}(b). Figs. \ref{fig6}(e) and \ref{fig6}(f) shows the $I_{\mbox{\scriptsize B}}$ dependence of $I_{\mbox{\scriptsize C}}$ as \emph{b}=0 and \emph{b}=-1, respectively. The former indeed predicts a parabolic dependence as observed in Figs. \ref{fig4}(c) and \ref{fig4}(d) while the latter also predicts a linear dependence besides of the parabolic one, which qualitatively reproduces the results in Figs. \ref{fig5}(c) and \ref{fig5}(d). It is worth bearing that field-like torque and damping-like torque are both indispensable to realize the asymmetry reduction of $I_{\mbox{\scriptsize C}}$ under opposite $I_{\mbox{\scriptsize B}}$. Fig. \ref{fig5} also indirectly manifests that the two types of torque both play important roles in magnetization switching process of the TFM system.

Other Pt/Co/MgO and Ta/CoFeB/MgO samples have exhibited similar switching symmetries. Noteworthy, though we demonstrate the switching behaviors with aid of an applied field, the switching performance controlled by $I_{\mbox{\scriptsize B}}$ will be still achievable in principle if the applied field is replaced by an effective field from exchange coupling.
\section{SUMMARY}
Current induced torques of Pt and $\alpha$-Ta, including damping-like torque and field-like torque, have been characterized by second-harmonic technique as $\beta_{\mbox{\scriptsize L, Pt}}$=-40 nm, $\beta_{\mbox{\scriptsize L, Ta}}$=+4 nm, $\beta_{\mbox{\scriptsize T, Pt}}$=+1.2 nm and $\beta_{\mbox{\scriptsize T, Ta}}$=-4 nm. Current can generate much larger field-like torque in $\alpha$-Ta than in Pt. Current-induced magnetization switching has also been realized in the $\alpha$-Ta system, indicating its high enough spin-orbit coupling strength and shedding light on its potential use in spin-orbitronics. Field-like torque, though incapable of switching \textbf{M} directly in our case, plays crucial role in reducing $I_{\mbox{\scriptsize C}}$.

$I_{\mbox{\scriptsize B}}$ results in different influences on switching behaviors for the TFM and PCM systems. Opposite $I_{\mbox{\scriptsize B}}$ equally decreases $I_{\mbox{\scriptsize C}}$ in PCM while it asymmetrically influences the $I_{\mbox{\scriptsize C}}$ in TFM system. Furthermore this asymmetry originates from the field-like torque of $I_{\mbox{\scriptsize B}}$ and can be adjusted by polarity of $H_{\mbox{\footnotesize \emph{y}}}$. Our work not only brings to light the influence of damping-like and field-like torques of switching current and bias current on switching but also experimentally demonstrates an electrical manner (via bias current) to symmetrically or asymmetrically control the switching, which could advance the development of spin-logic applications in which control of the switching process via electrical methods is crucial and beneficial.

\begin{acknowledgments}
This work was supported by the 863 Plan Project of Ministry of Science and Technology (MOST) (Grant No. 2014AA032904), the MOST National Key Scientific Instrument and Equipment Development Projects [Grant No. 2011YQ120053], the National Natural Science Foundation of China (NSFC) [Grant No. 11434014, 51229101, 11404382] and the Strategic Priority Research Program (B) of the Chinese Academy of Sciences (CAS) [Grant No. XDB07030200].
\end{acknowledgments}

\appendix*
\section{DETAILS OF MACROSPIN MODEL}
The schematic structure of the Pt/Co/MgO or Ta/CoFeB/MgO is shown in Fig. \ref{fig1}(a). An applied field $H_{\mbox{\footnotesize \emph{y}}}$ and the switching current (\emph{I}) are along the +\emph{y} axis. The bias current ($I_{\mbox{\scriptsize B}}$) is along the +\emph{x} axis. The ratio of $I_{\mbox{\scriptsize B}}$/\emph{I} is defined as a parameter \emph{c} which actually reflects the angle between the direction of total current density with that of the applied field. Easy axis of the perpendicular systems (PCM or TFM) is along the \emph{z} axis. Therefore the total energy (\emph{E}) is $K\sin^2\theta-\mu_0M_0H_{\mbox{\footnotesize \emph{y}}}\sin\theta\sin\varphi$  with \emph{K} anisotropy energy, $M_0$ saturation magnetization and $\mu_0$ permeability of vacuum. This energy drives an effective field \textbf{H}$_{\mbox{\footnotesize eff}}$=$-\nabla_{_{\mbox{\scriptsize M}}}$\emph{E}. Here we use a macrospin model for simplicity and therefore only $\theta$ and $\varphi$ are variable with the $M_0$ being a constant. $H_{\mbox{\footnotesize $\theta$,eff}}$=$-H_{\mbox{\footnotesize an}}\sin\theta\cos\theta+H_{\mbox{\footnotesize \emph{y}}}\cos\theta\sin\varphi$ and $H_{\mbox{\footnotesize $\varphi$,eff}}$=$H_{\mbox{\footnotesize \emph{y}}}\cos\varphi$. $H_{\mbox{\footnotesize an}}\equiv 2K/\mu_0M_0$. $H_{\mbox{\footnotesize $\theta$,eff}}$ and $H_{\mbox{\footnotesize $\varphi$,eff}}$ are two orthogonal components of \textbf{H}$_{\mbox{\footnotesize eff}}$. As the currents \emph{I} and $I_{\mbox{\scriptsize B}}$ are both applied, magnetization direction will be modulated due to the damping-like and field-like torques originated from the \emph{I} and $I_{\mbox{\scriptsize B}}$. The damping-like torque of a unit of \textbf{M} induced by the \emph{I} via spin Hall effect is defined as a parameter \emph{a} which is proportional to the spin Hall angle and along the \emph{x} axis. Then the damping-like torque induced by the $I_{\mbox{\scriptsize B}}$ is \emph{ac} which is however along the \emph{y} axis. As shown in the maintext, $\beta_{\mbox{\scriptsize L(T)}}$ is defined as the effective field correspond to the damping (field)-like torque induced by an unit of \emph{I}. Here we further define \emph{b} as $\beta_{\mbox{\scriptsize T}}$/$\beta_{\mbox{\scriptsize L}}$. Thus field-like torque induced by the \emph{I} via Rashba effect as well as Ostered mechanism is \emph{ab} and along the \emph{y} axis. In contrast, the field-like torque induced by the $I_{\mbox{\scriptsize B}}$ is \emph{abc} and along the \emph{x} axis. It is very important that the direction of the field-like torque induced by the \emph{I} is the same as that of the damping-like torque induced by the $I_{\mbox{\scriptsize B}}$. They are both along the \emph{y} axis. The final state of the system is determined by the following LLG equation (\ref{LLG}).
%\begin{widetext}
\begin{equation}\label{LLG}
\begin{aligned}
  -\frac{1}{\gamma}\frac{d\vec{M}}{M_0dt}=&-\alpha \vec{M}\times\frac{d\vec{M}}{M_0dt}+\frac{\vec{M}}{M_0}\times \vec{H}_{eff}\\
  &+a\frac{\vec{M}}{M_0}\times(-\hat{e}_x)\times \frac{\vec{M}}{M_0}+ab\frac{\vec{M}}{M_0}\times(-\hat{e}_x)\\
  &+ac\frac{\vec{M}}{M_0}\times \hat{e}_y \times \frac{\vec{M}}{M_0}+abc\frac{\vec{M}}{M_0}\times \hat{e}_y
\end{aligned}
\end{equation}
%\end{widetext}

In the first line of Equation (\ref{LLG}), $\gamma$ and $\alpha$ are gyromagnetic ratio and damping constant, respectively. The quantity \textbf{\emph{e}}$_{\mbox{\footnotesize \emph{x}}}$ and \textbf{\emph{e}}$_{\mbox{\footnotesize \emph{y}}}$ is unit vector along the \emph{x} and \emph{y} axis, respectively. The 1$^{\mbox{\footnotesize st}}$ and 2$^{\mbox{\footnotesize nd}}$ term in the second line is damping-like and field-like torque induced by the switching current (\emph{I}), respectively. The 1$^{\mbox{\footnotesize st}}$ and 2$^{\mbox{\footnotesize nd}}$ term in the third line is damping-like and field-like torque induced by the bias current ($I_{\mbox{\scriptsize B}}$), respectively. At the steady state, $d$\textbf{M}$/M_0dt$=$0$. Thus we arrive at Equation (\ref{App-torque}).
%\begin{widetext}
\begin{equation}\label{App-torque}
\begin{aligned}
  0=&\vec{m}\times \vec{H}_{eff}\\
  &+a\vec{m}\times(-\hat{e}_x)\times \vec{m}+ab\vec{m}\times(-\hat{e}_x)\\
  &+ac\vec{m}\times \hat{e}_y \times \vec{m}+abc\vec{m}\times \hat{e}_y
\end{aligned}
\end{equation}
%\end{widetext}
Here we have replaced \textbf{M}/$M_0$ with \textbf{m}. Equation (\ref{App-torque}) gives the scalar equations (\ref{App-simp}) which is also shown in the main text.
\begin{widetext}
\begin{subequations}\label{App-simp}
\begin{equation}
  H_y\cos\varphi-a\cos\theta\cos\varphi-ab\sin\varphi+ac\cos\theta\sin\varphi-abc\cos\varphi=0
\end{equation}
\begin{equation}
  H_y\cos\theta\sin^2\varphi-H_{an}\sin\theta\cos\theta\sin\varphi-a\sin^2\varphi+ab\cos\theta\sin\varphi\cos\varphi-acsin\varphi\cos\varphi-abc\cos\theta\sin^2\varphi=0
\end{equation}
\end{subequations}
\end{widetext}
As \emph{c}=\emph{b}=0, Equation (\ref{App-simp}) is reduced as Equation (\ref{App-simper})
\begin{widetext}
\begin{subequations}\label{App-simper}
\begin{equation}
  (H_y-a\cos\theta)\cos\varphi=0
\end{equation}
\begin{equation}
  sin\varphi(H_y\cos\theta\sin\varphi - H_{an}\sin\theta\cos\theta\ - a\sin\varphi)=0
\end{equation}
\end{subequations}
\end{widetext}
One possible solution as well as the final physically meaningful solution of Equation (\ref{App-simper}) is further reduced as Equation (\ref{App-simpest})
\begin{widetext}
\begin{subequations}\label{App-simpest}
\begin{equation}
  \cos\varphi=0
\end{equation}
\begin{equation}\label{App-simpest-b}
  H_{an}\sin\theta\cos\theta\ - H_y\cos\theta\sin\varphi + a\sin\varphi=0
\end{equation}
\end{subequations}
\end{widetext}
This solution shares the similar form with that derived in Ref.[\onlinecite{LiuPRL2012}], which demonstrates the rationality of our derivations.

In general case, Equation (\ref{App-simp}) can be transformed as Equation (\ref{App-general}).
\begin{widetext}
\begin{subequations}\label{App-general}
\begin{equation}
  \cos\varphi=\frac{(ab-ac\cos\theta)\sin\varphi}{H_y-a\cos\theta-abc}
\end{equation}
\begin{equation}\label{App-general-b}
  H_{an}\sin\theta\cos\theta-\frac{[(H_y-abc)^2+a^2(b^2+c^2+1)]}{H_y-a\cos\theta-abc}\cos\theta\sin\varphi+\frac{aH_y(1+\cos^2\theta)}{H_y-a\cos\theta-abc}\sin\varphi=0
\end{equation}
\end{subequations}
\end{widetext}
Comparing Equation (\ref{App-simpest-b}) and (\ref{App-general-b}), we find that the introduction of $I_{\mbox{\scriptsize B}}$ actually updates the $H_{\mbox{\footnotesize \emph{y}}}$ with an effective field of $[(H_{\mbox{\footnotesize \emph{y}}}-abc)^2+a^2(b^2+c^2+1)]/(H_{\mbox{\footnotesize \emph{y}}}-a\cos\theta-abc)$ and updates the \emph{a} with an effective torque of $aH_{\mbox{\footnotesize y}}(1+\cos^2\theta)/(H_{\mbox{\footnotesize \emph{y}}}-a\cos\theta-abc)$.

As \emph{c}=0, the effective field becomes $[H_{\mbox{\footnotesize \emph{y}}}^2+a^2(b^2+1)]/(H_{\mbox{\footnotesize \emph{y}}}-a\cos\theta)$. A nonzero \emph{b} can make the effective field larger, which is very beneficial for higher efficient switching. As \emph{b}=0, the effective field becomes $[H_{\mbox{\footnotesize \emph{y}}}^2+a^2(c^2+1)]/(H_{\mbox{\footnotesize \emph{y}}}-a\cos\theta)$. Therefore, the introduction of the bias current (or nonzero \emph{c} regardless of its polarity) can also increase the effective field. Besides, the field-like torque of the switching current (\emph{ab}) in the former case functions a similar role with the damping-like torque of the bias current (\emph{ac}) in the latter case. Only as \emph{b}$\neq$0 can \emph{c} with opposite sign asymmetrically influence the effective field. It is also worth noting that the $H_{\mbox{\footnotesize \emph{y}}}$ is still indispensable for magnetization switching because a zero $H_{\mbox{\footnotesize \emph{y}}}$ will also lead to a zero effective torque. The numerical results regarding the solutions of Equation (\ref{App-simp}) are shown in the Fig. \ref{fig6} in the main text and not shown here.

%\bibliography{references}

%

\end{document}